# Understanding Generative AI-mediated User Engagement with Academic Library Resources

Hae Min Kim is Senior Data Analyst, Libraries, Drexel University, email hk433@drexel.edu, ORCID 0000-0002-9128-0495 and Stacy Stanislaw is Director, Communications, Libraries, Drexel University, email svs22@drexel.edu, ORCID 0009-0006-4587-3966

## Abstract

This study empirically analyzed generative AI as an emerging discovery pathway to academic library resources. Utilizing web analytics from August 2023 to October 2025, the research identifies a significant increase in AI-mediated traffic, particularly following the integration of linked citation features. Referral analysis identified ChatGPT, Perplexity, and Gemini as the primary platforms driving this traffic. A substantial portion of users reached the institutional repository, primarily accessing electronic theses and dissertations. This pattern suggests that AI retrieval mechanisms effectively surface resources with structured metadata and stable permalinks that are Open Access and freely available. The results illustrate how AI ecosystems currently expose library resources and underscore the need for continued analysis and a strategic response to the evolving AI landscape.

## Introduction

Recent surveys indicate that nearly 90% of U.S. college students use generative artificial intelligence (AI) platforms such as ChatGPT to support their coursework, research, or study activities (Legatt, 2025). For students and researchers, these tools have become the new starting point for exploring unfamiliar topics, synthesizing information, and locating resources (Nasr et al., 2025). This widespread integration of AI into academic workflows represents a profound shift in how information is sought and encountered in higher education. As users increasingly begin their inquiries within AI interfaces, library resources are becoming part of this AI-mediated discovery environment.

Open access publications, instructional library guides, and resources stored in institutional repositories – traditionally accessed through search engines or library websites – are now being included in AI-generated responses. When users follow these links, they reach library resources through algorithmic recommendations or contextual exposure, rather than through deliberate searching.

As these new discovery pathways emerge, the integrity of AI-generated outputs depends heavily on their connection to the verifiable, authoritative sources that underpin them. As Kambhamettu et al. (2025) demonstrate, direct links to primary materials enable users to mitigate the risks of algorithmic oversimplification or hallucination. In this context, library-curated content serves as a critical reference framework, providing the authoritative evidence necessary for responsible and evidence-based AI use. Despite this potential, structural uncertainties constrain the seamless integration of library resources into AI ecosystems. A 2024 NISO Open Discovery Initiative survey highlighted a significant policy gap regarding the indexing of institutional repositories by AI crawlers (Varnum et al., 2025). While libraries anticipate improved visibility, they remain concerned about metadata transparency and the extractive nature of harvesting without reciprocal coordination, which threatens the trust and sustainability of library infrastructures.

These developments raise important questions about how library-curated content enters AI ecosystems and how users discover it. Despite emerging concerns and the heightened visibility of library resources in AI-generated responses, little is known about the scale or nature of this phenomenon. How frequently do users reach library resources through generative AI platforms? Which types of content are most visible in AI-generated responses, and how might these new discovery patterns differ from traditional search behaviors?

This study seeks to provide an initial empirical foundation for understanding AI-mediated information behavior and the emerging ways in which users encounter and engage with library resources through generative AI systems. By examining traffic sources from AI platforms using Google Analytics (GA4) data across multiple library service systems, this research aims to illuminate how AI mediation is reshaping the digital pathways that connect users to library-curated information resources.

Specifically, the study seeks to answer the following research questions:

**[A] Identification of AI-Mediated Access**

RQ1. Which generative AI platforms can be identified as referral sources in GA4 data?

RQ2. When did AI-mediated traffic first emerge, and how has it evolved over time?

RQ3. What proportion of total website access is mediated by AI-generated referrals?

**[B] Distribution Across Library Systems**

RQ4. To which library systems (e.g., core library website, library guides, discovery service, institutional research repository) do AI-mediated access lead?

RQ5. How do engagement metrics (e.g., user engagement, bounce rate) differ between AI-mediated and traditional referral traffic?

**[C] Characteristics of Accessed Content**

RQ6. What types of content (e.g., open access publications, subject guides, instructional materials) are most frequently accessed through AI-mediated exposure?

By analyzing GA4 data from an academic library, this research provides one of the earliest empirical accounts of AI-mediated information behavior. Moving beyond theoretical speculation, the findings characterize the scale and nature of algorithmic referrals, offering actionable insights for metadata optimization and digital service management.

Ultimately, this study will help inform how libraries can proactively refine their strategies for improving the discovery experience and maintain institutional relevance in an AI-driven information landscape.

## Literature Review

### *AI & Academic Libraries: Perceptions, Adoption, and Practices*

Perceptions of AI within academic libraries have evolved rapidly since the public emergence of generative AI tools, particularly following the release of ChatGPT to the public in late 2022. Early studies reflected uncertainty and limited preparedness, but more recent research indicates a gradual shift toward more positive and practice-oriented views of AI adoption. By 2025, the literature increasingly characterizes AI not as a speculative technology, but as a persistent component of the academic information environment that libraries must actively engage with through policy development, service design, and information literacy initiatives.

Survey-based studies of librarians provide insight into this transitional period. A comparative study of North American librarians in 2019 and 2025 revealed a growing gap between personal satisfaction and institutional reality (Peekhaus, 2026). While personal AI use has surged, institutional ownership remains low at 30.8%, and over a third of librarians are uncertain if their libraries even use AI. Interestingly, despite 75% reporting workplace interactions with AI, actual application in specific tasks like cataloging remains minimal, and only 25% feel confident teaching algorithmic literacy. Ultimately, with only 40% believing AI could replace aspects of their roles, librarians view the technology as a complementary tool that requires further professional training rather than a total replacement of expertise.

This pattern of cautious institutional adoption is consistent with other recent findings in the field. Lo (2024a), based on a U.S. nationwide survey conducted in 2023, found that academic library staff reported relatively low levels of AI literacy and limited confidence in applying AI tools to professional work. The study emphasized the need for structured training and proposed an AI literacy framework tailored to library context, incorporating technical understanding, ethical awareness, and critical evaluation skills.

Similarly, Castillo and Kelly (2025) surveyed library instructors and found that, while respondents generally perceived generative AI tools such as ChatGPT as useful, their intention to adopt these tools in instruction remained cautious. Importantly, the willingness to integrate AI into information literacy programs increased as student and faculty use of AI tools became more widespread, suggesting that user behavior itself is a key driver of librarian engagement.

Beyond studies of perception, empirical research has explored the integration of AI into academic library practices. Michalaka, Tzoc, and Lewis (2025) documented library-led adoption of generative AI tools at three mid-sized university libraries, emphasizing responsible implementation through attention to contracts and data privacy, bias mitigation, user education, local usage analysis, and continuous feedback collection. Case studies in instruction indicate that librarians are integrating generative AI into information literacy teaching. Johnson et al. (2024) described librarian-designed and librarian-delivered classroom activities for college students and faculty that used ChatGPT-generated texts as instructional materials. These activities enabled participants to practice critical evaluation, source verification using library resources, and known-item searching, while offering actionable guidance for implementing AI literacy instruction in academic libraries. Lo (2024b) similarly reported that targeted AI literacy instruction led to measurable improvements in college library users' AI-related skills and confidence, reinforcing the role of librarians as educators rather than mere technology adopters.

Within the broader literature on libraries' AI adoption, research has examined the application of generative AI in reference services. These studies investigate how AI tools are incorporated into operational workflows, assessing their capacity to augment professional practice while highlighting the continuing need for human judgment and contextual expertise. Yang and Mason (2024), comparing ChatGPT responses with those of reference librarians, found that librarians significantly outperformed the AI system in accuracy, relevance, and interpersonal quality. The study concluded that while generative AI has value as a support mechanism, it lacks the contextual and institutional awareness necessary for fully effective reference service.

These findings have prompted further investigation into how AI can be operationalized within library reference services, particularly in contexts where systems are intentionally designed to support, rather than replace, professional practice. Accordingly, subsequent research has moved beyond general-purpose AI tools to examine service-level integrations with reference services tailored to library-specific contexts. Chen and Chang (2025) developed and deployed an AI-enhanced reference framework that improved retrieval speed and response structure. But it remained dependent on human expertise for interpretive and culturally sensitive inquiries. These cases position AI as a complementary tool that augments, rather than supplants, professional judgment.

### *AI Use by Students and Researchers: Information Practices, AI Literacy, and Perceptions*

Recent literature consistently shows that generative AI tools, particularly ChatGPT, are already embedded in the everyday academic practices of college students and researchers. Across disciplinary and national context, students report using AI tools for a wide range of educational purposes, including assignment support, drafting and revising text, summarization, brainstorming, and search assistance. Rather than representing a speculative or emerging behavior, AI use has become a normalized component of learning and research workflows within a short period following the public release of generative AI tools.

Survey-based studies indicate that perceived usefulness is the strongest determinant of students' intention to use generative AI in higher education. Drawing on the Technology Acceptance Model, Pham et al. (2025) found that U.S. college students' adoption of ChatGPT was driven primarily by its perceived utility, with social influence also playing a significant role. Peer norms and shared experiences were especially influential among students aged 18 – 25, suggesting that AI adoption is shaped not only by individual assessment of functionality but also by collective academic culture. These findings imply that institutional responses to AI use may benefit from peer-led training, visible success cases, and structured support rather than restrictive policies.

Studies examining perceived benefits further reinforce the instrumental value students associate with generative AI. Yeung et al. (2025), in a survey of undergraduate and postgraduate students in Hong Kong, reported that more than 80% of respondents used generative AI for learning or research. Students viewed AI as effective in improving efficiency, supporting idea generation, and enabling personalized learning. Notably, undergraduate students reported higher levels of curiosity and exploratory learning than postgraduates, suggesting that AI may play different cognitive and motivational roles across academic stages. Similar positive perceptions of academic support were observed among U.S. college students, including across gender, race, and socioeconomic groups, although adopters consistently rated benefits more favorably than non-users (Zhang et al., 2025).

Despite widespread adoption and generally positive evaluations, the literature also documents persistent concerns related to accuracy, bias, and ethical use. Studies across disciplinary contexts report high rates of misinformation encounters. For example, Sultan et al. (2025) found that while over 90% of nursing students had used ChatGPT for academic purposes, nearly 87% reported experiencing inaccurate or misleading outputs. Although many students reported verifying AI-generated information, the frequency of errors underscores the limitations of uncritical reliance on AI tools, particularly in high-stakes domains such as healthcare education. Similar concerns regarding fabricated citations, hallucinations, and overreliance were reported in studies of management, pharmacy, and general undergraduate populations (Suchanek & Kralova, 2025; Alsakaker et al., 2025; Baek et al., 2024).

In this context, research emphasizes the need for structured AI literacy frameworks that extend beyond technical proficiency. AI literacy is now conceptualized as encompassing critical evaluation of outputs, transparency in use, ethical awareness, and an understanding of system limitations and bias. Multiple studies argue that prohibiting AI use is neither effective nor desirable. Instead, educators and librarians are encouraged to integrate AI into curricula with clear guidelines, reflective practices, and explicit discussion of risks and responsibilities (Rajabi et al., 2024; Suchanek & Kralova, 2025). These findings align with broader calls to move beyond binary debates about acceptance versus rejection and toward responsible, equitable, and pedagogically informed integration of AI in higher education.

Importantly, the literature suggests that AI-mediated practices reflect a shift in information behavior that differs from traditional models of information seeking. Rather than initiating deliberate searches within library systems, students often turn to AI tools as conversational intermediaries that generate synthesized responses, suggest sources, and shape problem formulation. Large-scale usage data further indicate that a substantial proportion of AI interactions among young adults involve “doing” tasks such as writing, summarizing, or producing content, rather than purely seeking information (Chatterji et al., 2025). This blurring of search, synthesis, and production challenges established distinctions between information retrieval and information use, raising new questions about how users encounter, evaluate, and attribute academic sources.

While user satisfaction and perceived usefulness are generally high, the literature consistently highlights gaps in critical evaluation skills, uneven ethical awareness, and uncertainty about appropriate uses. These limitations are important for understanding AI-mediated access to library resources, as they indicate that users may increasingly discover scholarly content not through deliberate engagement with library systems but through AI-driven exposure embedded within broader academic workflows. This shift necessitates an examination of the mechanisms by how AI platforms function as the primary gatekeeper of academic information.

### *AI as a Discovery Intermediary*

The studies describe generative AI systems as intermediaries in the discovery process, shaping how information is presented, interpreted, and accessed across fields. Rather than operating as neutral tools that simply retrieve user-specified results, AI systems actively mediate discovery through processes of summarization, recommendation, and automated linking. This shift broadens how people find information, with AI-mediated discovery now operating alongside traditional, user-driven search methods.

Cui and van Esch (2025) describe this transformation through the concept of "algorithmic fidelity," arguing that users often accept AI-generated outputs as authoritative despite limited transparency into how sources are selected or prioritized. In this model, users often encounter information passively rather than through deliberate searching. Ma et al. (2025) further demonstrate that AI systems have begun to operate as interpretive agents, shaping not only what content is accessed but also how it is framed and contextualized for users. These patterns indicate that discovery is shaped not only by user intent but also by algorithmic decisions built into AI systems. Research across multiple contexts reinforces the generality of this phenomenon. Kambhamettu et al. (2025) show that AI-driven systems consistently favor content that is structurally and semantically legible to machines, privileging standardized formats, clear metadata, and concise textual organization. This pattern appears across scholarly communication, media platforms, education, and e-commerce, suggesting that AI-mediated access is widespread and not limited to library environments.

Recent studies also examine the technical and semantic factors that influence which sources AI systems highlight, cite, or link. Work on retrieval-augmented generation (RAG) models emphasizes the importance of metadata quality, document clarity, and contextual signals (Ma et al., 2025; Mombaerts et al., 2024). Content lacking clear structure or contextual cues is less likely to appear in AI-generated responses, regardless of its intellectual value. Other structural factors also affect which materials become more visible. Kuratomi et al. (2025) demonstrate that openly accessible and well-linked materials are more likely to be retrieved, while Spennemann (2023) cautions that poorly structured sources increase the risk of fabricated or distorted citations. These findings show that AI-mediated discovery is not neutral but shaped by the technical features and limitations of these systems.

Within this environment, library-managed resources occupy a distinctive but understudied position. Institutional repositories, open-access publications, and electronic theses and dissertations often contain rich metadata and standardized formats that AI systems prefer. However, increased algorithmic visibility does not guarantee meaningful access. The NISO Open Discovery Initiative survey (Varnum et al., 2025) reflects growing concern among libraries and publishers regarding transparency, attribution, and governance in AI-driven discovery. Respondents note that improved visibility does not necessarily ensure access, as AI systems may present citations or summaries without providing clear links to authorized full-text content. Additional research identifies tensions emerging from this imbalance. Orduña-Malea et al. (2024) describe metadata as public infrastructure that supports AI-mediated discovery but is often harvested without transparency or reciprocity. Industry reports show that AI features are being embedded rapidly into discovery platforms, often outpacing institutional governance.

Taken together, this literature shows that AI-mediated discovery already shapes how information is encountered, linked, and valued. While previous research has examined perceptions of AI in libraries or patterns of AI use, fewer studies have traced how AI systems actually connect users to library-managed content. Addressing this gap, the present study analyzes where and how AI-mediated discovery directs users to library resources.

## Methodology

An exploratory case study was conducted aiming to discover patterns of AI-mediated access to online resources provided by the Drexel University Libraries. Web analytics data was extracted from Google

Analytics (GA4), which tracks user interactions across multiple library-managed platforms, including the library catalog, institutional repository, and core website. The dataset covers the period from August 1, 2023, through October 31, 2025. This timeframe was chosen to capture the initial emergence and growth of referral traffic originating from generative AI platforms.

### *Identification of AI Referral Sources*

To identify traffic mediated by generative AI platforms, the current analysis examined the "session source" dimension in GA4, which records the origin of user visits based on referring URLs.

In this study, generative AI platforms are defined as web-based or app-based tools primarily powered by large language models (LLMs) that generate new content such as text, summaries, or synthesized explanations in response to user prompts. This includes stand-alone conversational and multimodal interfaces such as ChatGPT, Microsoft Copilot, Google Gemini, Perplexity, Claude, and similar services. These platforms are distinct from AI-powered research tools or AI-enhanced discovery systems such as databases, search engines, or discovery layers that incorporate AI features for ranking, recommendation, or metadata enrichment, because those systems do not generate new textual or multimodal outputs but rather augment traditional retrieval processes.

Traffic was classified as "AI-mediated" when the referral source domain matched one of these LLM-based generative platforms. This filtering isolates sessions where access to library resources is likely initiated or facilitated through AI-generated responses or citations.

AI-mediated referrals were identified through a two-step process:

> (1) direct matching against known AI domains such as chatgpt.com, chat.openai.com, perplexity.ai, gemini.google.com/app, copilot.microsoft.com; and
>
> (2) pattern-based filtering to capture emerging or lesser-known sources using domains ending in ".ai" or containing AI-related identifies.

It is important to note that referral information from generative AI platforms is not always captured consistently across access modes, and the varying rollout dates of platform signals may lead to a conservative estimate of AI-mediated traffic.

### *Scope of Analyzed Library Managed Platforms*

Drexel University is a global R1 research institution serving 20,868 undergraduate and graduate students and 5,970 faculty and staff. Drexel's libraries are maintained to support the University's teaching and research mission by providing access to over 1.5 million online resources, including e-books, journals, databases, streaming videos, and curated research and course guides, through several different platforms. Access to these electronic resources is generally restricted to current students, faculty, and staff in order to comply with license agreements for educational and research use. The Libraries are also responsible for maintaining the University's institutional repository and facilitating access to Drexel's 14,416 electronic theses and dissertations and 104,631 research and scholarly works.

The analysis examined the following Drexel University Libraries systems and tools used to provide access to information resources:

- Library core website (Sitecore): The primary portal for service information, news & events, policies, and key navigation links to other library systems and services (www.library.drexel.edu).
- Library Discovery Service (ExLibris Primo): The central discovery layer for integrated searching of the Libraries' collections (drexel.primo.exlibrisgroup.com).
- Drexel Research Discovery Repository (ExLibris Esploro): The University's institutional repository hosts Drexel's research and scholarly outputs created by faculty, staff and students, including articles, theses and dissertations, datasets, and creative works (researchdiscovery.drexel.edu).
- Library Guides (Springshare): A platform for subject and course-specific research guides and curated instructional materials created by Drexel librarians (libguides.library.drexel.edu).
- Virtual Reference (Springshare): A knowledge base and virtual reference service where users access FAQs and submit questions to librarians (drexel.libanswers.com).
- Room & Event Scheduling (Springshare): The system for booking group study rooms, spaces, workshops, and appointments (libcal.library.drexel.edu).
- Digital Exhibits (Omeka): A platform for curated collections and special digital exhibitions created by the Drexel University Archives (drexelexhibits.omeka.net).

*Metrics and Dimensions*

GA4 provides multiple engagement metrics, including sessions, users, and views that represent distinct dimensions of online interaction. For this study, the primary unit of analysis is "total users," rather than sessions or views. This decision was made based on both institutional analytics experience and prior research indicating that bot crawlers and automated scripts frequently inflate session counts through repetitive access to library resources (Casden et al., 2025). Since such automated activities rarely maintain persistent identifiers across sessions, user counts provide a more reliable estimate of actual human engagement. By focusing on user-level data, this approach aims to reduce the distortion caused by automated traffic and better reflect authentic usage patterns. However, for content-level analysis, "sessions" were utilized to capture the frequency of access events, acknowledging that a single user may access multiple resources across different interactions. Additionally, supplementary indicators such as average engagement time, bounce rate, and page views were used to provide further context regarding user behavior.

Key GA4 dimensions analyzed include session source to identify referral origins, page path to show the specific library content accessed, and hostname to identify library systems. Collectively, these dimensions enable a comparative assessment of how different library systems are reached through generative AI referrals.

The analysis examined how AI-mediated access has evolved over time, highlighting shifts in the visibility and reach of library resources within generative AI environments. To answer the research questions, the study was structured into three phases:

(1) Monthly trend analysis to identify when and how traffic from generative AI platforms first appeared and changed over time;

(2) System-level distribution analysis to compare AI-mediated user traffic across major library systems to evaluate the relative impact of AI referrals on different service platforms; and

(3) Content-level analysis to identify the types of library resources most frequently accessed through AI-mediated referrals.

Collectively, these analyses aim to reveal emerging patterns of user interaction and resource discoverability shaped by the growing influence of generative AI platforms. Data processing and preliminary analysis were conducted in Microsoft Excel. Comparative analysis and visualization were performed using Microsoft Power BI to identify patterns and relationships within the dataset.

## Results & Discussion

### *Emergence and Growth of AI-Mediated Access*

The first stage of analysis examined referral sources recorded in GA4 to identify instances of traffic originating from generative AI platforms. This step aimed to determine the overall extent and diversity of AI-mediated access and to establish a working list of platforms contributing to referral traffic.

Analysis of GA4 referral data between August 2023 and October 2025 identified 20 distinct generative AI platforms that directed users to Drexel University Libraries' online systems. These sources included major AI services such as ChatGPT, Gemini, Copilot, Perplexity, and Claude, as well as research and writing specialized platforms such as NotebookLM, Consensus, and Dimensions. Table 1 summarizes the identified platforms. Most of these platforms operate on large language model (LLM) architectures, which enable generative and conversational capabilities such as text synthesis and summarization. Core models include OpenAI's GPT series, Google's Gemini, Microsoft's Copilot (based on GPT), Anthropic's Claude, and Mistral, among others.

**TABLE 1**
**Distribution of AI-Mediated Users Across Library Systems (Total Users)**

| AI Platforms\ Library Systems | Core library | Discovery service | Research Repository | Library Guides | Virtual Reference | Scheduling | Digital Exhibit |
|---|---|---|---|---|---|---|---|
| **chatgpt.com** | 292 | 43 | 2,836 | 514 | 18 | 25 | 8 |
| **perplexity.ai** | 46 | 14 | 367 | 147 | 6 | 1 | - |
| **gemini.google.com** | 27 | 4 | 234 | 106 | 7 | 2 | - |
| **copilot.microsoft.com** | 11 | 13 | 25 | 1 | - | - | 1 |
| **blackbox.ai** | - | - | 33 | 6 | - | - | - |
| **claude.ai** | 1 | 4 | 3 | 6 | - | - | - |
| **consensus.app** | - | - | 6 | - | - | - | - |
| **notebooklm.google.com** | - | - | 6 | - | - | - | - |
| **chat.deepseek.com** | - | - | 3 | 2 | - | - | - |
| **pcw.dimensions.ai** | - | - | 4 | - | - | - | - |
| **iask.ai** | 1 | - | 2 | 1 | - | - | - |
| **app.txyz.ai** | - | - | 3 | - | - | - | - |
| **app.getconch.ai** | - | - | 2 | - | - | - | - |
| **felo.ai** | - | - | 2 | - | - | - | - |
| **exa.ai** | - | - | 1 | 1 | - | - | - |

| **grok.com** | - | 1 | 1 | - | - | - | - |
|---|---|---|---|---|---|---|---|
| **app.undermind.ai** | - | - | 1 | - | - | - | - |
| **quillbot.com** | - | - | 1 | - | - | - | - |
| **apps.abacus.ai** | - | - | - | 1 | - | - | - |
| **chat.mistral.ai** | - | - | - | 1 | - | - | - |

AI-mediated traffic to Drexel University Libraries online systems first appeared on August 25, 2023, originating from Perplexity, marking the earliest recorded instance of access in the data set. Subsequent traces emerged from ChatGPT in late 2023, followed by Copilot in early 2024 and iAsk, Dimensions, and Gemini by mid-2024, reflecting the gradual diversification of AI sources driving users to library systems (Figure 1). This initial traffic was enabled as hyperlinks may have appeared directly within the generated response text, allowing users to navigate immediately to the referenced page.

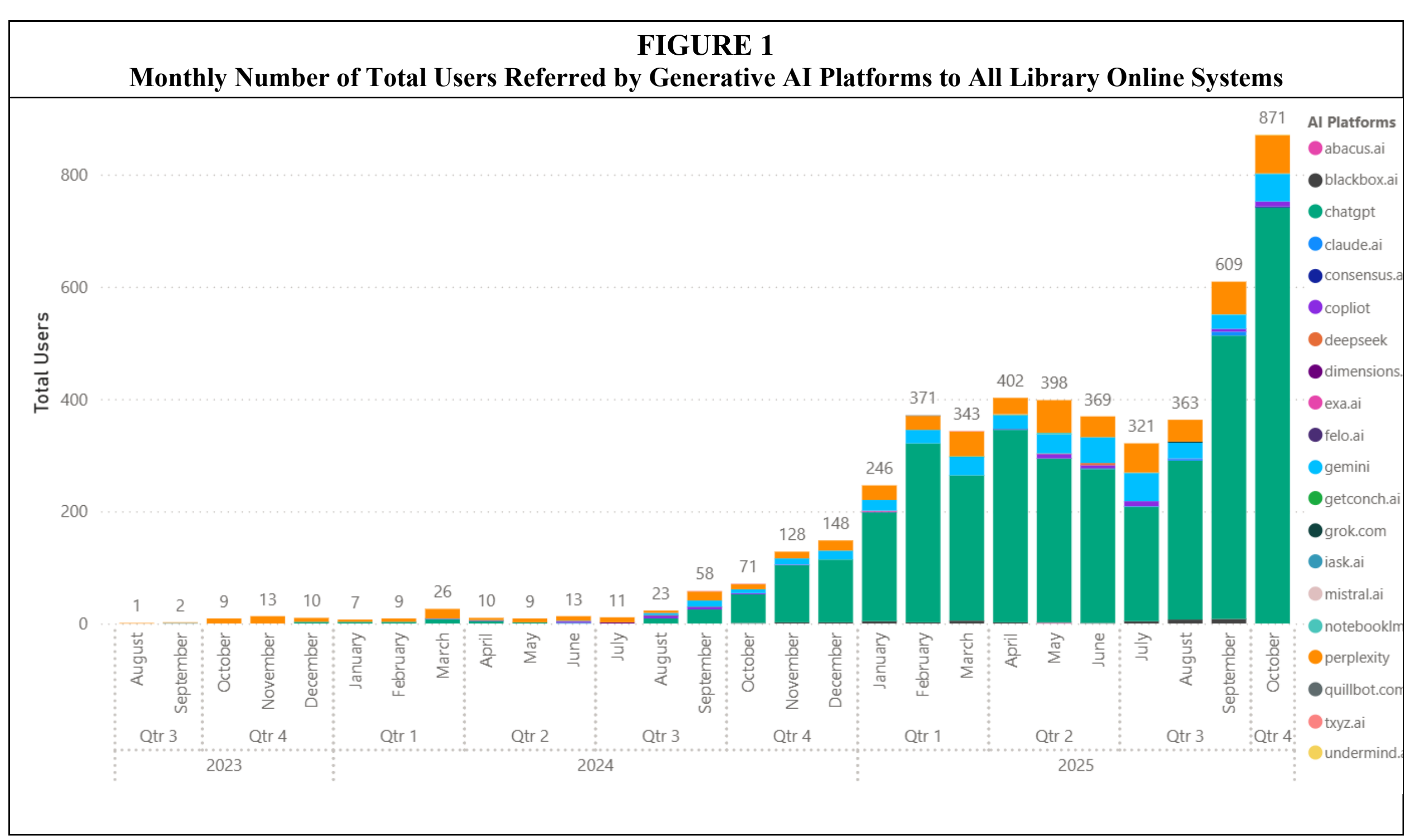

**FIGURE 1**
**Monthly Number of Total Users Referred by Generative AI Platforms to All Library Online Systems**

Among the AI platforms, ChatGPT has become the predominant referral source, showing a sustained growth pattern over time. Until October 2024, user traffic attributed to ChatGPT remained sporadic, whereas referrals from Perplexity, though small in volume, appeared more consistent over time. At the end of October 2024, OpenAI launched the “Sources” feature, enabling the model to display clickable reference links to external websites, including library resources (OpenAI, 2024; OpenAI, 2025). Unlike inline citations, which highlight only the most relevant references, the Sources list includes additional URLs that the model considered, potentially increasing pathways to access external content. The launch of the Sources feature coincided with a marked increase in traffic from ChatGPT, suggesting that enhanced visibility of verifiable sources may have enhanced user engagement with library domains.

Following this point, AI-mediated traffic experienced concurrent increases across multiple platforms. Perplexity, Gemini, and Copilot also demonstrated upward trajectories beginning in late 2024. This implies a collective trend of enhanced visibility and interoperability between generative AI outputs and library-based content. By October 2024, AI-mediated users reached their highest recorded level, 43% more than previous month. To contextualize this growth, Figure 2 illustrates the monthly proportion of AI-mediated access relative to total library online systems traffic. While AI-mediated referrals remain a minority of overall access, the figure shows a steady upward trend in their proportional share, indicating that AI platforms are becoming a more consistent source of referral traffic rather than a transient phenomenon.

**FIGURE 2**
**Monthly Proportion of AI-Mediated Access Relative to Total Traffic to All Library Online Systems**

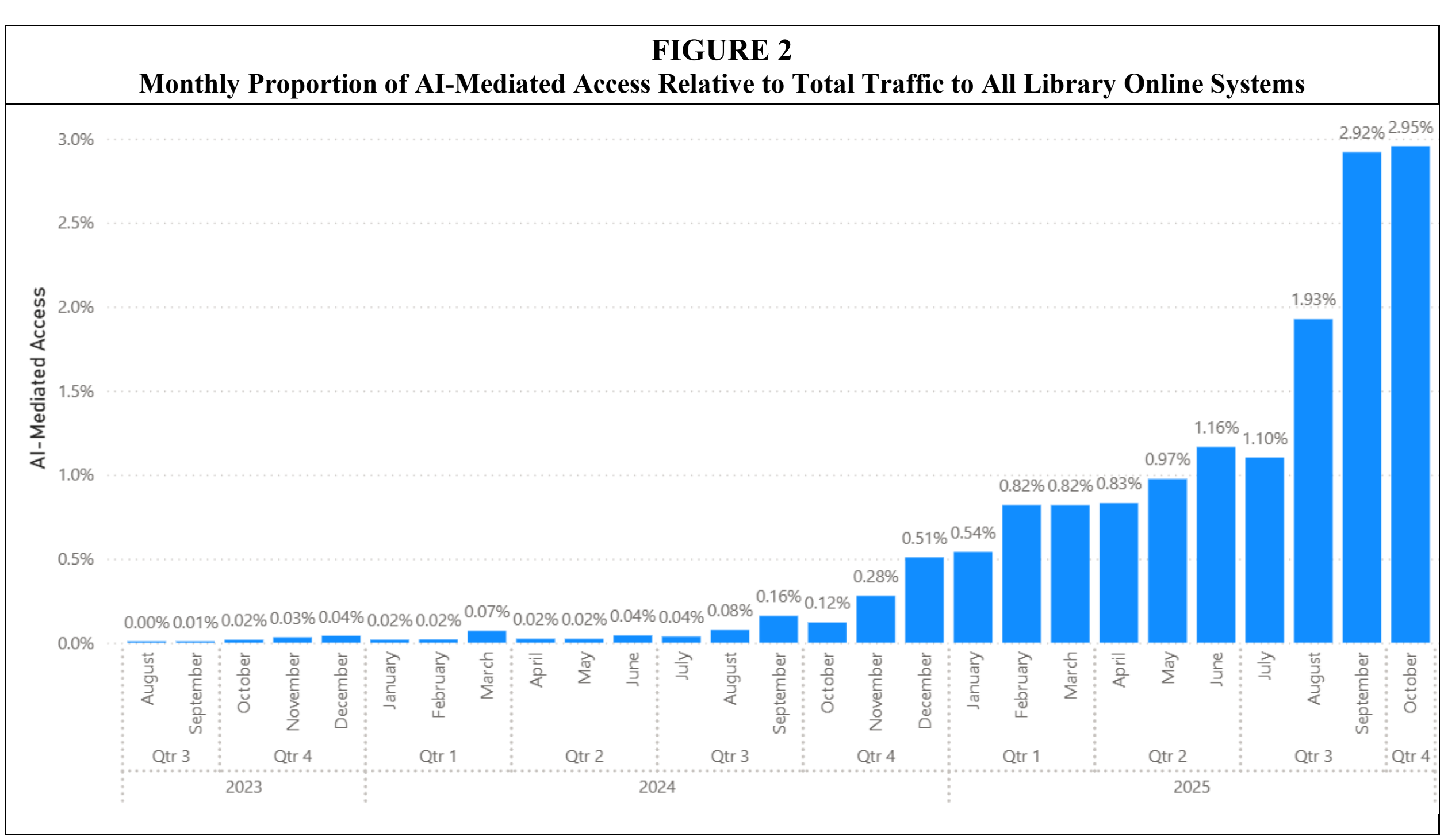

One possible factor that may help explain the sharp increase observed in 2025 is the evolution of generative AI platforms themselves. In particular, GPT-5's release in August 2025 introduced substantial improvements in model efficiency and factual reliability, with reports indicating notable reductions in hallucination rates and more consistent use of up-to-date information when web search is enabled (OpenAI, 2025b). This enhancement, combined with the introduction of the Sources feature in ChatGPT in late October 2024, expanded the frequency and visibility of outbound references generated in AI-produced responses. Following the launch of GPT-5, referrals via ChatGPT increased substantially. In September 2025, ChatGPT-mediated traffic rose by 78% compared to August, and in October 2025, it grew by 47% relative to the previous month. This suggests that improvements in model accuracy and citation behavior may have increased the likelihood that users encounter and subsequently follow AI-generated links to library resources. While this correlation does not establish causation, the alignment or technical advancements and traffic increases indicate that platform-level changes may have contributed to the observed patterns of AI-assisted discovery.

The distribution of AI-mediated referrals in October 2025 shows that ChatGPT had the largest share of traffic to the library website (85%), followed by Perplexity (8%), Gemini (6%), and Copilot (1%) (Figure 1). This pattern parallels broader global market trends reported, in which ChatGPT maintained an estimated 81% worldwide market share, with Perplexity (11%), Copilot (3.4%), and Gemini (3.0%) according to StatCounter (2025). While the absolute values differ, the relative ordering and concentration among the top platforms are similar, suggesting that local referral dynamic at the library may reflect broader patterns in user adoption and reliance on leading generative AI services.

### *Distribution Across Library Systems*

Analysis of GA4 referral data shows that AI-mediated access reached all seven major library systems during the observation period. This indicates broad exposure of Drexel University Libraries' digital platforms within generative AI environments. However, the distribution of this traffic was highly uneven. Research Repository (Esploro) received by far the largest number of AI-mediated users (2,997 users out of a total 159,940 repository visitors), followed by Library Guides (688 of 94,937) and the core library website (290 of 138,356).

Other systems saw only modest levels of AI-generated referrals, including Virtual Reference (29), Scheduling (17), and Digital Exhibit (7). The pronounced concentration of AI-mediated traffic in Research Repository aligns with the two major spikes in ChatGPT-driven referrals, "Sources" feature in November 2024 and GPT-5-release in October 2025, both of which corresponded with increases in Research Repository access. Figure 3 presents the monthly trend in AI-mediated users across library systems, showing that increases in overall AI-mediated access were driven primarily by growth in Research Repository with other systems remaining relatively stable.

Platform-specific patterns also differentiate how AI systems return library content. Of the 20 AI platforms observed, Research Repository received referrals from 18 distinct platforms, and Library Guides from 11, indicating that these two systems are more frequently indexed, cited, or linked across generative AI environments. This breadth of platform coverage contrasts with other systems, each of which received referrals from six or fewer AI platforms. This pattern suggests that subject information resources such as Library Guides and Research Repositories are more likely to appear in AI-generated results. In contrast, task-oriented or operational systems appear far less frequently. Notably, Discovery Service (Primo) received comparatively little AI-mediated traffic. This pattern suggests that generative AI platforms may be more likely to return item-level resources or subject specific guide content rather than directing users to broader discovery-layer entry points.

To examine differences in engagement patterns across referral types, this study focused exclusively on user interactions within the Research Repository that received the largest volume of AI-mediated traffic. Engagement metrics were analyzed for the most substantial referral sources: search engines (Google, Google Scholar, Bing); AI-mediated access via ChatGPT; and two institutionally integrated systems – Drexel Discovery Service and Drexel Learn, the learning management system (Blackboard Learn) – representing traditional, search-oriented access pathways.

**FIGURE 3**
**Monthly Trends in Total Users Across Library Systems**

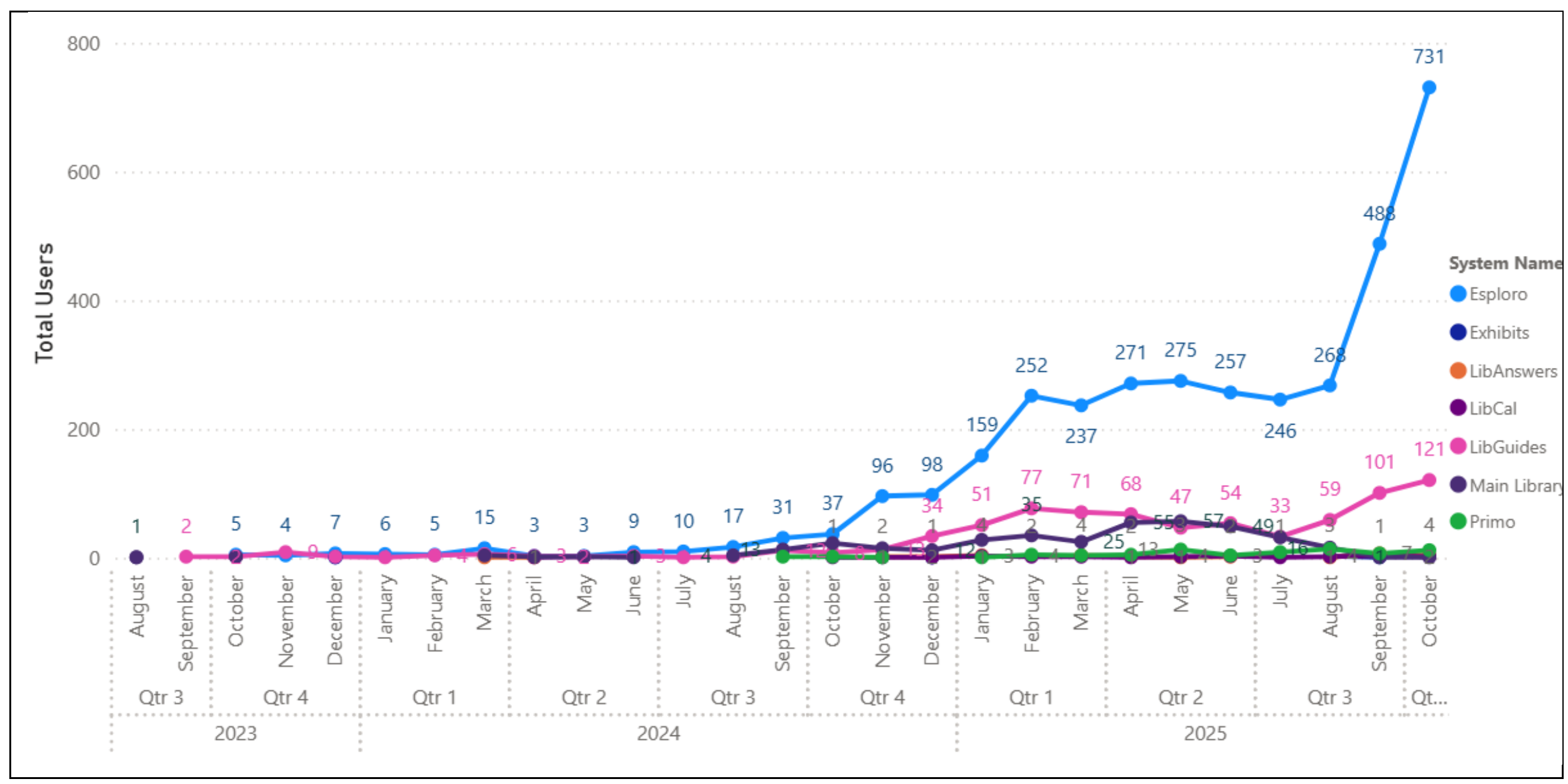

Across these sources, clear differences emerged in user behavior and session structure. Table 2 summarizes key performance indicators, including total users, user engagement, average engagement time per session, bounce rate, and views per active user during the observed period. Google accounted for the largest share of users and served as the primary entry point to the Research Repository. Despite the high volume, the average engagement time per session was relatively short (33.57 seconds), and the bounce rate remained moderate (26%). Google Scholar contributed fewer users but showed a higher bounce rate (46%) and lower engagement time (23.9 seconds), indicating more transactional visits. Bing generated a smaller user base but showed deeper engagement, with an average engagement time of 41.25 seconds and 3.44 views per active user. This suggests that users arriving via Bing were more likely to explore multiple items within the Research Repository. AI-mediated referrals from ChatGPT displayed moderate engagement (21.82 seconds) and a bounce rate of 42%. These metrics suggest a highly transactional and item-specific mode of interaction, where the user's engagement is focused on the landing page provided by the AI rather than navigating the broader library resources.

In contrast, referrals to Research Repository from Drexel Discovery Service and Drexel Learn demonstrated the strongest engagement metrics, including substantially longer average engagement time and significantly more views per active user. Users entering through these platforms are typically already engaged in structured discovery workflows or course-embedded research activities, making additional browsing or further navigation more likely. Their behavior suggests a continuation of an ongoing search process rather than a targeted, item-specific interaction.

Taken together, the engagement metrics show that ChatGPT-mediated traffic aligns more closely with search-engine referrals than with academic platform-driven visits. ChatGPT users tend to engage in brief, item-focused interactions, reflected in higher bounce rates, shorter durations, and limited within-site exploration. These patterns suggest that AI-mediated access appears to function more as a discrete and point-of-need reference pathway than as an entry point for broader discovery process.

| TABLE 2<br>Engagement Metrics across Referral Sources to Research Repository | | | | | | |
|---|---|---|---|---|---|---|
| **Type** | Session source | Total users | User engagement (Seconds) | Avg. engagement time per session (Seconds) | Bounce rate | Views per active user |
| **Search Engine** | google.com | 124,371 | 5,529,143 | 33.57 | 0.26 | 2.57 |
| | scholar.google.com | 17,582 | 488,420 | 23.98 | 0.46 | 1.40 |
| | bing.com | 2,942 | 181,124 | 41.25 | 0.16 | 3.44 |
| **Generative AI** | chatgpt.com | 2,396 | 71,975 | 21.82 | 0.42 | 1.86 |
| **Institutional Platforms** | drexel.primo.exlibrisgroup.com | 821 | 107,913 | 71.70 | 0.39 | 7.13 |
| | learn.dcollege.net | 574 | 139,917 | 127.20 | 0.24 | 9.47 |

### *Content-level patterns of AI-mediated access*

To identify the types of content most frequently accessed through AI-mediated exposure, we analyzed referral landing page URLs originating from AI platforms. Each URL was classified by (1) system of origin (e.g., core library websites, Discovery Service, Research Repository, Library Guides) and (2) material type (e.g., dissertations, books). The dataset contained 3,451 unique landing pages, which collectively received 9,753 sessions generated through AI referrals. Because the purpose of the research question was to assess the overall scale and distribution of AI-driven traffic, the analysis for this part focused on session volume to measure the frequency of content exposure. Based on session counts, Research Repository accounted for 75% of all referrals (7,303 sessions), followed by Library Guides (12%), the core library website (7%), and Discovery Service (4%). These proportions formed the basis for the subsequent content-type analysis.

AI-mediated referral analysis showed that Research Repository accounted for the majority of all incoming sessions. Because of this concentration, we conducted a detailed content-type analysis of Research Repository referrals. Research Repository URLs follow a consistent and interpretable structure, allowing content identification directly from the URL itself. Among the 7,303 Research Repository-based sessions, electronic dissertation and thesis (ETD) records dominated the distribution with 4,098 sessions (56%), followed by Journal articles with 1,430 sessions (20%) (Table 3). Library Guides followed, accounting for 1,217 sessions overall.

| TABLE 3<br>Research Repository Document Types | | |
|---|---|---|
| **Doc Type** | **Sessions** | **Percentage** |
| **Dissertation &thesis** | 4,098 | 56.1% |
| **Journal article** | 1,430 | 19.6% |
| **Search outputs** | 643 | 8.8% |
| **Book chapter** | 319 | 4.4% |
| **Profile** | 207 | 2.8% |
| **Conference** | 186 | 2.5% |
| **unknown** | 180 | 2.5% |
| **Book** | 90 | 1.2% |

| **Outputs** | 68 | 0.9% |
|---|---|---|
| **Report** | 35 | 0.5% |
| **Researchers** | 35 | 0.5% |
| **Abstract** | 12 | 0.2% |
| **Total** | 7,303 | 100.0% |

To better understand why Research Repository content, ETDs in particular, dominated AI-mediated referrals, we conducted an additional examination of technical factors influencing how AI systems identify and return web content.

Recent research and system documentation indicate that AI retrieval processes appear to work more effectively with web resources that are highly structured, semantically rich, consistently formatted, and publicly accessible. As a result, such resources may be more likely to surface in AI-generated responses. Contemporary AI tools such as ChatGPT utilize a retrieval-augmented generation (RAG) mechanism that enhances a model's output by adding external context during runtime (OpenAI, 2025c). Rather than depending on its pre-trained knowledge, RAG fetches relevant data from connected sources and incorporates it into the prompt, enabling the model to deliver responses that are more accurate and contextually informed. RAG-based workflows prioritize resources that can be reliably indexed, segmented, and ranked, making pages with clear semantic markup, predictable URL patterns, and well-exposed metadata highly advantageous for AI discovery (Microsoft, 2025; OpenAI, 2025a; OpenAI, 2024; OpenAI, 2025c; Google Cloud, 2025).

Under these retrieval mechanisms, machine-readable metadata (e.g., schema.org fields, controlled vocabularies) improves the ability of AI systems to understand a page's topical relevance and contextual boundaries (Mombaerts et al, 2024). Likewise, semantic page structures, such as consistent HTML hierarchy, embedded identifiers, and stable hyperlink relationships, increase a page's retrievability by enabling more accurate chunking, ranking, and citation during the generation process (OpenAI, 2025c). Recent documentation from major AI providers reinforces this trend: for example, OpenAI's description of the "sources" or "attribution" feature indicates that linkable, stable, and semantically coherent pages are more likely to be surfaced as explicit reference sources in generated responses (OpenAI, 2024). Similar behaviors have been documented across other retrieval-enhanced models, which consistently return content with clean metadata structure and high indexability (Li & Sinnamon, 2024).

Taken together, these developments suggest that AI systems do not treat all web resources equally. Instead, they tend to favor resources whose technical and semantic structures align with retrieval-oriented workflows. In this respect, Research Repository (Esploro) matches many of the characteristics that contemporary AI models prioritize. ETDs in Research Repository are typically single, self-contained scholarly objects with a clear title, abstract, and full text sequence and are open access. Each record is exposed through a stable static permalink, a simple and human-readable URL structure, and item-level landing pages enriched with structured metadata, including schema.org fields, controlled vocabularies, and persistent identifiers (Alter & Neuwirth, 2018; Ex Libris, 2025; Veltzman & Elstein, 2024). Full-text PDFs are hosted within the same domain, reinforcing both authority and indexability. These features collectively create a highly crawlable, machine-readable, and semantically coherent document profile.

While this alignment does not guarantee prioritization, it plausibly contributes to the prominence of Research Repository's ETDs within AI-mediated responses.

However, these alignments do not operate in isolation. AI referrals are also shaped by broader ecosystem factors, such as search engine indexing behavior, AI model tuning, web popularity signals, and question-dependent semantic matching. Thus, Research Repository's design appears to position its ETD records in a structurally advantageous place within AI retrieval workflows, although this advantage interacts with multiple contextual variables that can shift across platforms and use cases.

### *Behavioral and Strategic Implications*

Based on this study's data analysis, AI-mediated exposure appears to facilitate a more transactional mode of engagement rather than deeper or sustained exploration of library resources. Evidence from Google Analytics reveals that user behavior originating from ChatGPT-mediated referrals aligns more closely with general search engine traffic rather than institutionally embedded academic platforms. These interactions are primarily item-focused, characterized by brief session durations and high bounce rates, suggesting that users may utilize AI-generated links for item-specific inquires or quick verification rather than exploratory research.

In contrast, access originating from academically situated environments, such as Discovery Service or Blackboard Learn, demonstrates significantly more robust engagement metrics. These activities are characterized by longer durations and a higher volume of views per active user, reflecting the continuation of established research or learning process. While AI-mediated referrals represent discrete, instrumental access events, traditional discovery layers and course-integrated platforms support an ongoing search cycle where users are more likely to browse, compare, and navigate across multiple resources.

These different engagement patterns suggest that AI-mediated exposure can be understood as an emerging pattern of information behavior that differs from traditional search models. Classical frameworks, such as Kuhlthau's (1988) Information Search Process (ISP), emphasize intentional user-driven search and offer limited explanatory power when applied to AI-mediated contexts observed in this study. While the concept of "information encountering" (Erdelez, 1995) provides a useful lens for serendipitous discovery, AI-mediated exposure appears distinct in that moments of chance are no longer shaped primarily by environmental conditions, but by algorithmically engineered forms of serendipity. From this perspective, the notion of Algorithmic Fidelity (Cui & van Esch, 2025), offers a productive interpretive framework. Rather than positioning users as active seekers who initiate discovery, AI-mediated systems surface and contextualize information on the user's behalf.

As generative AI becomes more deeply embedded in academic workflows, the strategic importance of AI-mediated visibility is likely to increase. Recent studies underscore a widespread adoption of these tools among college students and researchers (Legatt, 2025; Chatterji et al., 2025), alongside the growing presence of institutional AI literacy initiatives and governance frameworks (Michalaka et al., 2025; Varnum et al., 2025). As these technologies normalize, the frequency that users engage with library resources through AI-generated responses will likely increase, necessitating a shift in library discovery strategies.

This study suggests that AI-mediated access has emerged as an indirect yet distinctive discovery pathway that complements traditional intent-driven searching. Within this framework, library resources are increasingly surfaced through algorithmic ranking and system-level synthesis rather than through explicit user queries. Because these systems seem to rely on structured representations to generate and rank responses, the visibility and discoverability of scholarly content in AI environments depend on the technical and semantic alignment of library metadata with evolving AI retrieval architectures.

To remain discoverable in this evolving landscape, library resources will need to align with the technical requirements of Retrieval-Augmented Generation (RAG) architectures. These systems prioritize content characterized by rigorous structure, semantic clarity, and high-quality metadata (Kambhamettu et al., 2025; Ma et al., 2025). While institutional repositories often meet these criteria, broader library-managed assets may require significant optimization to achieve parity in AI-mediated discovery (Varnum et al., 2025). This discrepancy highlights a critical need to evolve metadata practices beyond internal system compliance toward enhanced machine readability. Essential technical interventions include the adoption of standardized schema, stabilization of persistent identifiers (URLs), and the alignment of content structures with formats optimized for AI processing (Cui & van Esch, 2025; Kambhamettu et al., 2025; Ma et al., 2025; Mombaerts et al., 2024).

These efforts should not be viewed as responses to a temporary trend. Rather, they align with ongoing changes in how information is surfaced, interpreted, and accessed in digital environments. By refining metadata frameworks to be more AI-responsive and supplementing them with item-level summaries, libraries will not merely increase visibility; they will reinforce their core values of stewardship, transparency, and equitable access (Varnum et al., 2025). Ultimately, this proactive stance positions libraries to exert meaningful influence over how scholarly knowledge is encountered and utilized within the AI-mediated environments.

## Limitations

Despite providing an empirical foundation for understanding AI-mediated access, several technical and methodological limitations must be acknowledged.

First, the technical inconsistency of referral tracking likely results in a conservative estimate of AI-driven traffic. Referral information is not captured uniformly across all access modes. For example, mobile or desktop applications of generative AI tools may not transmit referral data in the same manner as browser-based sessions. Given the possibility of missed interactions, the observed data may underestimate the actual volume of AI-mediated engagement. Furthermore, because major platforms began transmitting identifiable referral signals at different times, the longitudinal trends are influenced not only by changes in user behavior but also by timing-related inconsistencies in platform data transmission.

Second, this study identified and classified AI sources based on specific criteria. We focused exclusively on platforms powered by large language models (LLMs) capable of generating new, synthesized content. Consequently, AI-powered research tools or embedded AI features within traditional discovery systems, which augment retrieval rather than generate text, were intentionally excluded. While the study used domain-level patterns such as ".ai" suffixes, and known platform lists to filter traffic, less prominent or newly emerging services may have been omitted.

Third, although GA4 metrics are inherently subject to automated activity, this study mitigated such distortions by prioritizing user-level data over session counts. This approach reduces the inflation caused by bot-generated sessions that lack persistent identifiers (Casden et al., 2025). Nevertheless, residual noise from non-human traffic may still exist within the dataset, representing a persistent challenge in web analytics research.

Finally, as a single-institution case study centered on Drexel University Libraries, the findings reflect a specific digital architecture and metadata environment. Variations in institutional SEO strategies, repository structures, and metadata richness could lead to different visibility patterns in other academic contexts. Moreover, while the correlation between platform updates (including the launch of GPT5 or the "Sources" feature) and traffic spikes is notable, it does not establish definitive causality, as seasonal academic cycles or internal literacy initiatives may have served as variables. Future longitudinal studies incorporating qualitative user data will be essential to further understand the evolving relationship between AI-mediated access and academic library resources.

## Conclusion

This study provides empirical evidence that researchers are already discovering and accessing library-managed resources through AI-mediated pathways, confirming the emergence of generative AI platforms as active intermediaries in scholarly discovery. Although this study is based on data from a single institution and AI-mediated access currently represents a small fraction of total library traffic, its trajectory suggests a notable change. The consistent growth observed over the past two years, coupled with the widespread adoption of generative AI among students and researchers, indicates that this discovery pathway is poised to expand as these tools become further embedded in academic and scholarly workflows.

As AI platforms become intermediaries in scholarly discovery, the strategic value of machine-readability appears to be a significant implication for library infrastructure. The concentration of traffic within the Research Repository (Esploro) suggests that AI retrievers such as RAG mechanisms may prioritize certain technical configurations. In particular, structured metadata, stable permalinks, and Open Access environments emerged as prominent factors associated with AI-mediated selection. From the behavioral data, brief session durations and high bounce rates in this study suggest a transactional engagement pattern. Users appear to utilize AI-mediated pathways for fact-checking or source verification in this study's data set. These findings suggest that machine-level discoverability may influence how library resources are represented and encountered within AI-mediated information environments, with implication for institutional visibility and the communication of authoritative content in an evolving AI-driven landscape.

Although AI has the potential to significantly improve information discovery, the challenges of AI should not be underestimated. As librarians, we should continue to monitor these technological changes, remaining current on AI advancements to effectively guide patrons in using AI tools for their teaching, learning and research. Also, the library community should engage in a conversation regarding the ethical and functional integration of its resources into AI environments. Just as other information sectors, such as journalism, are establishing stances on automated data harvesting, libraries should define our roles to ensure scholarly integrity is maintained.

Ultimately, future research should investigate AI-mediated access to library resources and seek to better understand how generative AI tools impact search behavior and information discovery. While this study provides a quantitative foundation by analyzing two and half years of access data, it does not capture the nuances of users' experiences behind these visits. Qualitative methods including user interviews or usability tests should be employed to investigate how AI-mediated access specifically benefits researchers and how it impacts their overall information discovery. This will provide critical insights for the future design of library services.